\documentclass[showpacs,floatfix,superscriptaddress,showpacs,twocolumn,amssymb,amsfonts,prb,aps]{revtex4}

\usepackage{longtable,graphicx,epsfig,dcolumn}

\begin{document}
\bibliographystyle{revtex}

\title{Crystalline Surface Phases of the Liquid Au-Si Eutectic Alloy}

\author{Oleg~G.~Shpyrko}
\altaffiliation[Current address: ]{Department of Physics,
University of California, San Diego,
La Jolla, CA 92093}
\affiliation{Department of Physics and School of Engineering and Applied Sciences,
Harvard University, Cambridge, MA 02138, USA}
\affiliation{Center for Nanoscale Materials, Argonne National Laboratory, Argonne, IL
60439, USA}
\author{Reinhard Streitel}
\affiliation{Department of Physics and School of Engineering and
Applied Sciences, Harvard University, Cambridge, MA 02138, USA}
\author{Venkatachalapathy S. K. Balagurusamy}
\affiliation{Department of Physics and School of Engineering and
Applied Sciences, Harvard University, Cambridge, MA 02138, USA}
\author{Alexei~Yu.~Grigoriev}
\affiliation{Department of Physics and School of Engineering and
Applied Sciences, Harvard University, Cambridge, MA 02138, USA}
\author{Moshe~Deutsch}
\affiliation{Department of Physics, Bar-Ilan University, Ramat-Gan
52900, Israel}
\author{Benjamin~M.~Ocko}
\affiliation{Condensed Matter Physics and Materials Science
Department, Brookhaven National Laboratory, Upton NY 11973, USA}
\author{Mati Meron}
\affiliation{Center for Advanced Radiation Sources, The University of
Chicago, Chicago, IL 60637, USA}
\author{Binhua Lin}
\affiliation{Center for Advanced Radiation Sources, The University of
Chicago, Chicago, IL 60637, USA}
\author{Peter S. Pershan}
\affiliation{Department of Physics and School of Engineering and
Applied Sciences, Harvard University, Cambridge, MA 02138, USA}

\date{\today}
\begin{abstract}
\def\baselinestretch{1}
\noindent A two dimensional crystalline layer is found at the
surface of the liquid eutectic Au$_{82}$Si$_{18}$ alloy above its
melting point $T_M=359~^{\circ}$C. Underlying this crystalline layer we find a
layered structure, 6-7 atomic layers thick.
This surface layer undergoes a first-order solid-solid phase transition
occurring at $371~^{\circ}$C. The crystalline
phase observed for T$>$371~$^{\circ}$C is stable up to at least
430~$^{\circ}$C. Grazing Incidence X-ray Diffraction
data at T$>$371~$^{\circ}$C imply lateral order comprising two
coexisting phases of different oblique unit cells, in stark
contrast with the single phase with a rectangular unit cell
found for low-temperature crystalline phase $359~^{\circ}$C$<T<371~^{\circ}$C.
\end{abstract}


\pacs{68.03.-g, 61.66.Dk, 68.35.Bs, 61.25.Mv}

\maketitle

\section{Introduction}

Liquid bulks exhibit short range order, extending to a few
molecular diameters only. In certain liquids, however, a free
surface induces ordering of the near-surface molecules into well-defined layers that typically extend into the bulk to depths of the order of the bulk correlation length.
For liquid crystals this can be several tens of
molecular lengths; however, for more typical atomic liquids layering
extends only 3 or 4 molecular diameters.\cite{Magnussen95,Regan95, Lei96, Lei97,
Regan97, Dimasi98, Tostmann98, Tostmann99, Yang99, Yang00,
Tostmann00, Dimasi00, Dimasi01, Huber02a, Huber02, Huber03, Li02,
Yang03, Shpyrko03,Shpyrko04, Li05, Shpyrko05,
Balagurusamy07} For all pure metals and non-dilute alloys, atoms
within each surface-induced layer were found to exhibit only
liquid-like short range order in the surface-parallel directions.

This general observation differs strikingly from our recent discovery
of a two-dimensional crystalline surface phase in the
Au$_{82}$Si$_{18}$ liquid eutectic alloy.\cite{Shpyrko06} Prior to this discovery
observation of laterally ordered phases in metallic liquids has been
limited to binary dilute systems at concentrations close to the phase boundary for
coexistence with the 3D solid crystal. In those cases, the dilute, higher melting point
component segregated to the surface to form a complete
surface monolayer.\cite{Yang99, Yang00, Yang03} The observation of
a surface-crystalline monolayer in a non-dilute system, which is far from the
phase boundaries of the bulk solid phase, is, therefore, quite unusual
and very unexpected. The formation of a crystalline surface phase
in Au$_{82}$Si$_{18}$ alloy is also accompanied by a surface layering enhancement of 6-7 well-defined atomic layers, manifested by an $\approx$30-fold enhancement of the
intensity of the Bragg-like layering peak in the
x-ray specular reflectivity, as compared to that observed for all
other metallic liquids measured to date.

At T=$371~^{\circ}$C the x-ray reflectivity curve was found to
undergo an abrupt change to a different profile, in which
the intensity of the reflectivity peak was reduced by a factor of
about 5, indicating a first-order surface phase transition. Here we
present evidence revealing that at $371~^{\circ}$C the quasi-2D
crystalline layer undergoes a phase transition into a different
crystalline surface phase.

Interest in the Au$_{82}$Si$_{18}$ eutectic alloy is due to a number of
factors. AuSi has an unusually deep eutectic point: typically the
melting point for the eutectic composition is lower by
100-200$~^{\circ}$C than the melting points of its individual
pure components. For AuSi the eutectic point (359$~^{\circ}$C) is many hundreds of degrees below
the melting points of Au
(1063$~^{\circ}$C) or Si (1412$~^{\circ}$C).\cite{Massalski}
The reasons for such an unusually deep eutectic point are still not fully
understood and attract, therefore, considerable attention.
Remarkably, AuSi also does not form thermodynamically stable crystalline intermetallic compounds for any temperature or composition and this has been speculated to be related to the deep eutectic point. Upon cooling below the eutectic
point, Au$_{82}$Si$_{18}$ either phase separates into pure Au
and pure Si regions, or, if quenched rapidly enough it forms an
amorphous solid. AuSi was the first discovered metallic glass\cite{Duwez60} and amorphous AuSi can be produced with a
variety of techniques, including splat-quenching and
evaporation.\cite{Weis92}

In addition to the fundamentally important questions of its unusual
bonding properties, AuSi eutectic has found a variety of practical
applications. Both Au and Si are crucial to the electronics industry, and thin gold wires are commonly used to interconnect silicon-based devices. The low melting point of the eutectic alloy, the eutectic's deep and narrow shape in the phase
diagram, and the lack of intermetallic crystalline compounds have
resulted in AuSi being used as a low-melting solder for Micro- and Nano-Electro-Mechanical Systems (MEMS and NEMS).\cite{Cohn96, Cheng00} The same properties are employed
to manufacture high-purity Si nanowires and other nanostructures
from AuSi alloys through a vapor-solid-liquid growth mechanism.
\cite{Wagner64,Plass97, Hu99, Hannon06, Kodambaka07} Finally,
extensive work has been done over the last 20 years on thin solid films of
AuSi, among other metal-semiconductor compounds, in connection with
Schottky barrier diode devices.

Despite of the large amount of research performed on solid bulk and thin
solid films of amorphous AuSi phases, remarkably little is known
about the properties of the liquid AuSi eutectic alloy, in
particular about atomic structure and the compositional variations in the
near-surface region. The Gibbs segregation rule predicts a
significant enhancement (65-85 at. {\%}) of Si in the surface
monolayer,\cite{Gibbs} which would result in a local composition
that is far from the eutectic composition of 18 at. {\%} of Si.
The high brightness of the x-ray beams generated by modern
synchrotron sources, in combination with the high-precision liquid surface reflectometers available at some at these sources,\cite{Lin03} make possible detailed, atomic resolution,
investigations of the near-surface structure of liquids. These
methods are employed here to determine the surface structure of the
AuSi eutectic over an extended temperature range from its melting
point up.

\section{Experiment}

X-ray measurements were carried out at ChemMatCARS, APS,
Argonne National Laboratory, and at beamline X22B, NSLS, Brookhaven
National Laboratory, using x-rays of a wavelengths
$\lambda=0.7743$~{\AA} and $1.5322$~{\AA}, respectively. A
number of measurements were carried out on each of the two different
samples used, yielding highly reproducible data.

Each AuSi sample was prepared from 100~g Au$_{82}$Si$_{18}$ ingots
of 99.99$\%$ purity (Goodfellow) melted under UHV
conditions (P$<10^{-9}$ Torr) in a Mo sample pan. The
resultant liquid sample had a diameter of 20~mm, a thickness of
$\sim$5~mm and a surface radius of curvature of $\sim$20~m.
Macroscopic surface oxides were scraped off by an in-vacuum Mo
scraper. Several hours of Ar$^{+}$ ion sputtering removed
the remaining microscopic oxide traces. The sample temperature was
electronically controlled to $\sim$0.01~$^\circ$C.
Pickup of acoustic noise and mechanical vibrations was
eliminated by placing the UHV sample chamber on an active vibration
isolation unit mounted on the liquid surface reflectometer.

It is important to describe the experimental steps that were
undertaken to guarantee that the signatures of the crystalline
surface phase observed in the x-ray diffraction and reflectivity
measurements are indeed due to the intrinsic surface structure of the AuSi alloy,
rather than being an experimental artifact, such as the formation of an oxide or
chemical impurities segregated at the surface. The presence of even
minute concentrations of surface impurities would have been detected by
surface-sensitive probes, such as x-ray photoemission spectroscopy (XPS) and
x-ray fluorescence (XF) measurements performed in grazing incidence
geometry. However, no signature of any impurity was, in fact,
detected by these techniques. Since these surface probes are capable
of detecting even a small fraction of a monolayer of another
species, the crystalline monolayer reported here is not likely to be
due to impurities. Reproducibility of the data for two AuSi samples
make an artifact due to contamination even less likely.

Additional evidence for lack of contamination or oxidation at the
surface comes from Ar$^+$ ion sputtering that was applied to the
surface during the experiments. Measurements performed
while sputtering, immediately after the end of the sputtering session
(typically lasting several hours), and a long time (24 hrs) after the
sputtering has ended, have produced identical and
highly reproducible results. Re-formation of a sputtered-off surface layer by surface
segregation of impurities from the bulk or oxidation of the
surface under UHV conditions (partial pressure of oxygen and water
was less than 10$^{-12}$ Torr according to in-situ residual gas
analysis of the chamber's contents) should occur within something of the
order of an hour (or more), which would have stood out clearly during the
series of measurements that were carried out.

Details of the x-ray scattering geometry of the experiments
have been previously described in detail.\cite{Tostmann99,
Shpyrko03, Shpyrko04a, Shpyrko_Thesis} The x-ray reflectivity,
$R(q_z)$, is the fraction of the intensity of the incident x-rays
reflected by a surface, for a grazing angle of incidence $\alpha$.
$q_z=(4\pi/\lambda)\sin \alpha$ is the surface-normal wavevector
transfer, and $q_c = 0.078$~{\AA}$^{-1}$ is the critical $q_z$ for
total external reflection in our case.\cite{AlsNielsen94}

For an ideal case of sharply terminated, static, structureless
interface, where the electron density profile normal to the
surface is a step function, reflectivity is described by the Fresnel law:
\begin{equation}
R_F(q_z)= {\biggl|
{{{q_z-\sqrt{{q_z}^2-{q_c}^2}}} \over
{q_z+\sqrt{{q_z}^2-{q_c}^2}}}
\biggr|}^2 \label{eq:Fresnel1}
\end{equation}
where $q_c\approx 4 \sqrt{\pi r_e \rho}$ is the critical
wavevector, $r_e= 2.813\cdot10^{-5}$~{\AA}\ is the classical
electron radius and $\rho$ is the electron density at the surface.
Reflectivity from a \emph{real} liquid surface deviates from
$R_F$ in two significant ways. First, the free surfaces are
never ideal, and exhibit a monotonic or a non-monotonic variation
of the electron density across the interface between the (constant)
value of the vapour and the (constant) value of the bulk. Second, at
any finite temperature the surface of every liquid is roughened by
thermally excited capillary waves. These waves partially destroy the
constructive interference between x-rays reflected from different
points on the surface, leading to a Debye-Waller-like reduction of
the specularly reflected intensities, the magnitude of which is
determined by the liquid's surface tension, temperature, and the
portion of the capillary waves' spectrum sampled by the
reflectometer.\cite{Sinha88, Pershan99, Pershan00}

For $q_z>4-5 q_c$, $R(q_z)$ is well approximated by:
\begin{equation}
R(q_z)=R_F(q_z) \cdot {\biggl| \Phi (q_z) \biggr|} ^2 \cdot CW(q, T,
\gamma), \label{eq:Fresnel2}
\end{equation}
where $CW$  is a surface roughness term due to the
capillary waves \cite{Sinha88}, and $\Phi(q_z)$ is the
surface structure factor, given by:\cite{Braslau88}
\begin{equation} \Phi (q_z) = \frac{1}{\rho_{\infty}} \int
\textrm{d}z\frac{\textrm{d} \langle \rho(z) \rangle }{\textrm{d}z}
\exp(\imath q_z z). \label{eq:structure}
\end{equation}
Since $R_F$ is a universal function, depending only on the measured
$q_c$, and $CW$ is obtained from measurements of the off-specular
diffuse scattering,\cite{Tostmann99,Shpyrko03,Shpyrko04a} $\Phi$ can be determined, within a phase factor, \cite{AlsNielsen94} from the measured $R(q_z)$ through
Eq.~(\ref{eq:Fresnel2}). This, in turn, allows a determination of
the laterally-averaged surface-normal electron density profile
$\langle \rho(z) \rangle $ from Eq.~(\ref{eq:structure}) by a fit to a physically-motivated model for $\rho(z)$.

For liquid metals a surface-induced layering is observed.
Similar to Bragg diffraction from a crystal, constructive
interference between x-rays reflected from these well-defined
surface-induced atomic layers leads to a significant enhancement of
the reflected signal at $q_z\approx2\pi/a$ where $a\approx 3$~{\AA} is the spacing
of the atomic layers. In the case of the standard layering observed
for most liquid metals, the electron density is typically described
by a semi-infinite sum of equidistantly spaced gaussians, normalized
to the bulk density $\rho _{\infty }$:
\begin{equation}
    \frac{\langle \rho(z) \rangle}{\rho _\infty} = \sum _{n=0} ^\infty
    \frac{d/\sigma _n}{\sqrt{2\pi }}
    \exp \left[ -(z-nd)^{2}/2\sigma _n ^2 \right]
\label{eq:rho}
\end{equation}
The widths of the gaussians are defined by $\sigma _n ^2 = n\bar{\sigma}^2 + \sigma _{0}^{2}$, where $\sigma _0$ and $\bar{\sigma}$ are adjustable fit parameters.
This form for $\sigma _n$ produces a gradual increase in the
Gaussian width with distance $z$ below the surface, and a corresponding decrease in their height. Since they are equally spaced, this leads to their
gradual merging, yielding eventually a constant density for large $z$.
Thus, $\bar{\sigma}$ is inversely related to the decay length for
the surface layering. In all liquid metals studied to date the typical layering extends to 3-4 well-defined layers before density oscillations decay to $\rho_{\infty}$.

The density profile that had to be used to fit the measured AuSi reflectivity curves included N additional layer-representing Gaussians (where usually N=4 or 5),  with adjustable densities ($\rho_i$), positions ($x_i$) and widths ($\sigma_i$):
\begin{eqnarray}
 \frac{\langle \rho(z) \rangle}{\rho _\infty}=
    \sum_{i=1}^N \rho_i \exp \left[ -(z-z_i)^{2}/2\sigma _i ^2 \right]+ \nonumber\\
    + \frac{d/\sigma _n}{\sqrt{2\pi}}\sum _{n=0} ^\infty
    \exp \left[ -(z-z_N-nd)^{2}/2\sigma _n ^2 \right]
\label{eq:rho_ausi}
\end{eqnarray}
Finally, the properties of the thermally excited capillary waves are governed by
the competition between the thermal energy, described by the temperature
$T$ and the restoring force, provided by the surface tension $\gamma$.
Capillary waves induce surface roughness that gives rise to diffuse
scattering away from the specular condition, thus reducing $R(q_z)$.
Detailed analysis of the lineshape of
the off-specular diffuse scattering intensity allows one to
determine the term $CW(q, T, \gamma)$ in Eq.~\ref{eq:Fresnel2}.\cite{Tostmann99, Shpyrko03,
Shpyrko04a} In practice any specular reflectivity measurement
involves integration of the off-specular diffuse scattering
contributions over the finite resolution function of the detector.
Details of the capillary wave theory and the resolution effects on capillary
contribution are described elsewhere.\cite{Sinha88, Tostmann99,
Shpyrko03, Shpyrko04a}
\begin{figure}[t]
\includegraphics[angle=0,width=1.0\columnwidth]{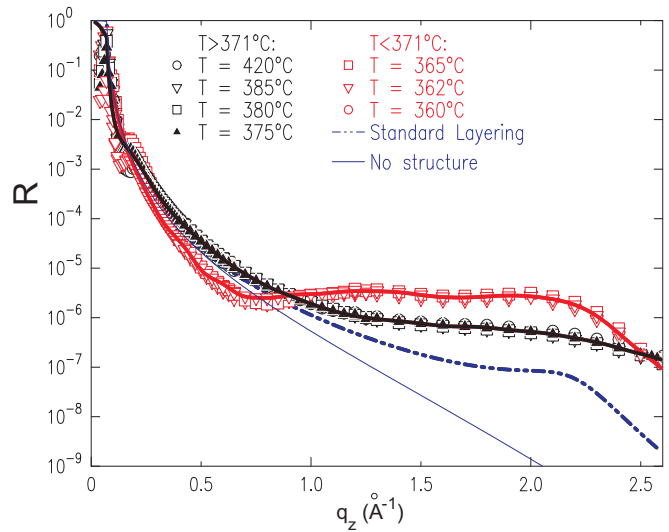}
\caption {Measured x-ray reflectivity data for T$<$371~$^{\circ}$C (red symbols) and T$>$371~$^{\circ}$C (black symbols) along with their
corresponding model fits (lines) discussed in the text.
The reflectivity expected for a standard 3-4 surface
layers, observed in all other metallic liquids, is shown in a dashed blue line.
The solid blue line shows the Fresnel reflectivity curve, expected
for an ideally flat and smooth liquid AuSi alloy's surface.}
\label{fig:ausi_fig1}
\end{figure}

\section{Results}

\subsection{X-ray Reflectivity}

\begin{figure}[t]
\vspace{1mm} \centering
\includegraphics[angle=0,width=1.0\columnwidth]{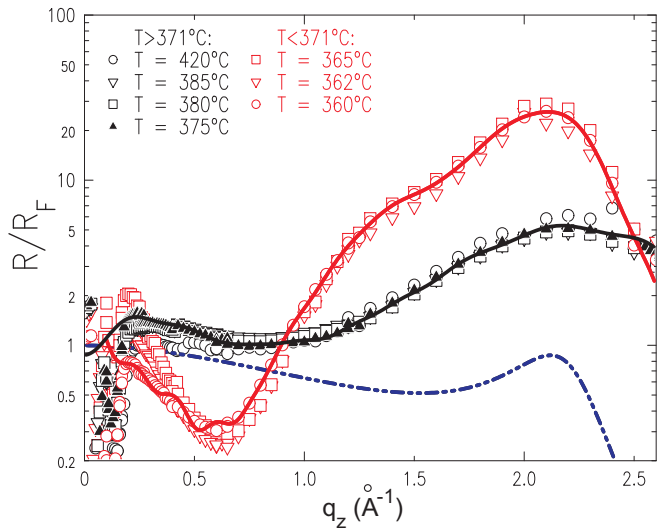}
\caption{Fresnel-normalized reflectivity $R/R_F(q_z)$ from the liquid AuSi
surface for different temperatures. Notation as in
Fig.~\ref{fig:ausi_fig2}. Note the logarithmic $R/R_F$ scale.} \label{fig:ausi_fig2}
\end{figure}

The as-measured and the Fresnel-nomalized x-ray reflectivity curves
are shown at five different temperatures in Fig.~\ref{fig:ausi_fig1} and Fig.~\ref{fig:ausi_fig2}, respectively.
Two features stand out immediately. The first feature is the bimodal temperature dependence, with a sharp division between the shapes of the low-temperature (T$<$371~$^{\circ}$C, red symbols) and the high-temperature (T$>$371~$^{\circ}$C, black symbols) $R/R_F$, but with practically no shape variation within each
temperature range. The second feature is that $R/R_F$ in \emph{both} regimes are
much higher than that expected on the basis of the conventional layering
that was observed in almost all liquid metals and alloys studied to date. The low-$T$
$R/R_F$, in particular, exhibits an $\approx$30-fold increase over the standard
layering peak. The transition temperature between the two regimes,
$T_0=371.15$~$^{\circ}$C, was determined by measuring $R/R_F$
at a fixed $q_z$ while varying $T$, and found to be hysteresis-free
to within our $\pm 0.1$~$^{\circ}$C
$T$-resolution~\cite{Shpyrko06}.

 \begin{figure}[b]
\vspace{1mm} \centering
\includegraphics[angle=0,width=1.0\columnwidth]{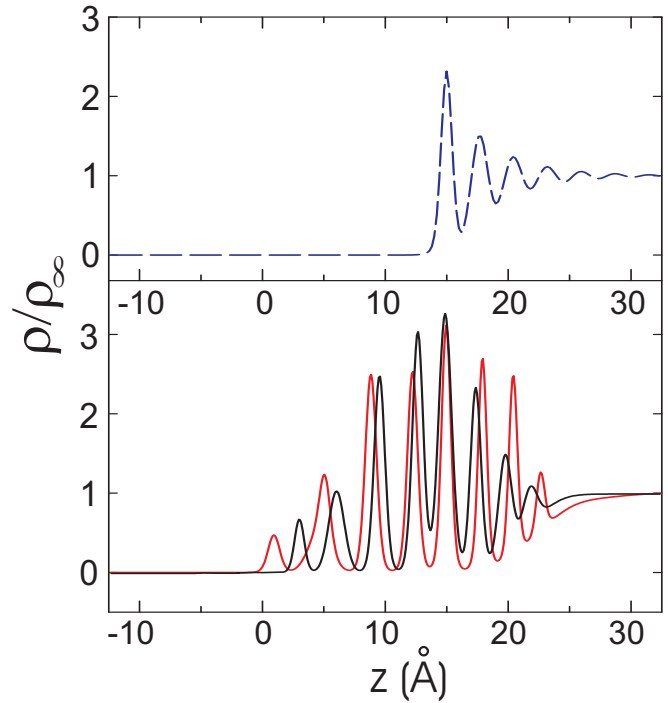}
\caption{Density profiles for a standard layering model (blue dashed
line, upper panel), and for the low-temperature (red line) and high-temperature
(black line) phases of the liquid AuSi surface (lower panel). The three density
profiles shown here were obtained from the fits of the x-ray reflectivity
curves shown in the same-color lines in
Figs.~\ref{fig:ausi_fig1} and \ref{fig:ausi_fig2}.}
\label{fig:ausi_fig3}
\end{figure}

The model fits shown in lines in Figs.~\ref{fig:ausi_fig1} and \ref{fig:ausi_fig2},
yield the electron density profiles plotted in
Fig.~\ref{fig:ausi_fig3}, along with that expected from conventional
layering in AuSi. It should be noted that some specific details of
the density profile cannot be determined uniquely from such
fits, due primarily to the relatively large number of the adjustable parameters in using density model in Eq.~\ref{eq:rho_ausi}.
This model includes N=4-5 adjustable layers in addition to 3-4 layers employed in fitting conventional surface layering.
Specifically, the precise positions, relative intensities and
widths of the layers are correlated, albeit weakly, and fits of
similar quality could be obtained by correlated variations in the values
of several of these parameters within restricted ranges of
values. Some of these ambiguities can be lifted by taking into account the
physical details of the density model - for example, a layer
cannot be narrower than the diameter of the atom, and the density
can not rise above that of a close-packed layer of atoms. Despite this uncertainty,
the main physical features of the fits, such as the existence of at
least 6-7 well-defined, atomically sharp, almost periodically ordered layers are essential for
reproducing the measured $R/R_F$ shown in
Figs.~\ref{fig:ausi_fig1} and \ref{fig:ausi_fig2}. These features are, therefore,
strongly supported by the measured curves. As Fig.~\ref{fig:ausi_fig3} demonstrates, both low- and high-$T$ density profiles show a similarly well-ordered
surface-normal structure. However, the high-$T$ profile
exhibits a higher degree of disorder, reflected in fewer and lower
layering peaks in Fig.~\ref{fig:ausi_fig3}. The increased
disorder in the high-$T$ phase results in a lower intensity of the
corresponding quasi-Bragg layering peak of the high-$T$ phase $R/R_F$ curves in
Fig.~\ref{fig:ausi_fig2} as compared to that of the low-$T$ phase $R/R_F$.
Both low- and high-$T$ layering peaks in Fig.~\ref{fig:ausi_fig2} are significantly
broader than the conventional layering peak, shown in a dash-dot-dot line. This
effect is likely due to a slight aperiodicity of the
layering, that is also apparent in the density profiles in
Fig.~\ref{fig:ausi_fig3}.

As the electron densities of Au and of Si are, respectively,
much higher and much lower than the average density of the alloy,
$\rho_{\infty}$, the $\rho/\rho_{\infty}<1$ relative density of the
top surface layer in Fig.~\ref{fig:ausi_fig3} implies a Si
enrichment relative to the bulk's composition. Using the fitted density profile and assuming closed-packed
homogeneous layers of Au and Si atoms of known diameters, a simple
calculation yields a $\sim70$ atomic {\%} of Si atoms in the topmost
surface layer, in good agreement with the 65-85 atomic {\%} of Si
predicted by the Gibbs adsorption rule for an ideal binary solution of
AuSi.\cite{Gibbs}

\subsection{Grazing-incidence diffraction}

\begin{figure}[t]
\vspace{1mm} \centering
\includegraphics[angle=0,width=1.0\columnwidth]{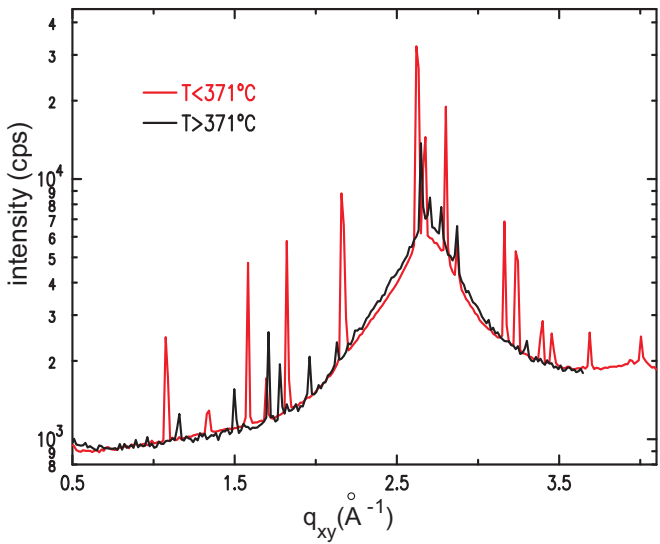}
\caption{Grazing Incidence Diffraction patterns for low temperature
(red line) and high temperature (black line) surface phases.
The sharp peaks indicate the presence of a
crystalline surface phase. The broad underlying peak is the
liquid bulk scattering peak.}
\label{fig:ausi_fig4}
\end{figure}

The surface-parallel ordering was explored by
grazing-incidence X-ray diffraction (GIXD).\cite{Dosch92,AlsNielsen94, Daillant00,Dutta00} X-rays incident on the surface below the critical angle $\alpha_c$ generate an evanescent wave which travels along the surface, and has an
amplitude which decays exponentially with depth below the surface,
with a 1/e decay length of $\Lambda \approx 1/q_c = 14$~{\AA}. Thus,
the x-rays sample only the top few atomic layers. Scattering by
surface-parallel order present in these layers gives rise to GIXD patterns. Such
patterns measured for the low-$T$ and the high-$T$ phases are shown in
Fig.~\ref{fig:ausi_fig4}. In addition to the broad peak arising from
the short-range order of the underlying bulk liquid, one can
clearly see a number of sharp diffraction peaks indicating existence
of long-range in-plane order. This is the first, and so far
only, liquid metal or non-dilute alloy to show surface-parallel
order. All previous GIXD measurements on liquid metals and
non-dilute alloys revealed only a broad liquid-like peak,
indicating that despite the increased order in the
surface-normal direction, i.e. the surface-induced layering, the
order within each layer in the surface-parallel directions remained
liquid-like.

As observed in Fig.~\ref{fig:ausi_fig4}, the GIXD patterns
found above and below $T_0\approx 371$~$^{\circ}$C are not
identical, implying a change in the lateral order of the surface.
This confirms that the sharp structural transition
detected in the reflectivity measurements, discussed above, is
indeed a first-order solid-solid surface phase transition. We now
discus the order in the two phases, as revealed by the GIXD
measurements.

Indexing of the GIXD pattern for the low-$T$
surface phase has been previously reported\cite{Shpyrko06}
within a 2D rectangular $7.386 \times 9.386$~{\AA}$^2$ unit
cell, with a Au$_4$Si$_8$ stoichiometry. This composition is in
quantitative agreement with the electron density derived from the
$R/R_F$ fit. The unit cell dimensions, which determine the positions of the peaks observed
in the GIXD pattern, were determined using the CRYSFIRE
\cite{Crysfire} powder indexing software. The automatic exhaustive
search for the crystallographic structure allowed indexing all
observed peaks in a single two-dimensional unit cell. The simplest
Bravais lattice symmetry was rectangular, and attempts to fit the
data by an oblique Bravais lattice invariably resulted in a
primitive unit cell with $\gamma=90.0\pm0.3^{\circ}$. A nearly perfect match was obtained for 22 Bragg peaks that were within our measurable range of GIXD measurements, even though some peaks (for example (23) and (31), or (22) and (13) pairs) could not be resolved with resolution provided by the soller slit setup. Positions of
atoms within the unit cell were refined using the GSAS,
\cite{GSAS} PowderCell \cite{PowderCell} and CaRine \cite{CaRine}
software packages. The initial composition was varied from pure Au to
pure Si, including Au$_4$Si, Au$_2$Si, AuSi, AuSi$_2$
stoichiometries. While during the diffraction pattern refinement the
number of atoms within the unit cell was defined roughly by the
requirement of close packing, in principle it is also conceivable
that various atoms of quasi-2D crystalline structure are displaced
up or down, and are not strictly in the same horizontal (liquid
surface-parallel) plane. In fact, such "puckering" may also be one of
the reasons why formation of crystalline 2D layer does not result in
the nucleation of a 3D bulk phase. The
distribution of the atoms along the surface-normal direction can be
obtained, in principle, from a detailed analysis
of the intensity distribution along the Bragg rods
of the different GIXD peaks\cite{Robinson92}, a complete measurement
of which has not been done in the present study.

\begin{figure}[b]
\vspace{1mm} \centering
\includegraphics[angle=0,width=1.0\columnwidth]{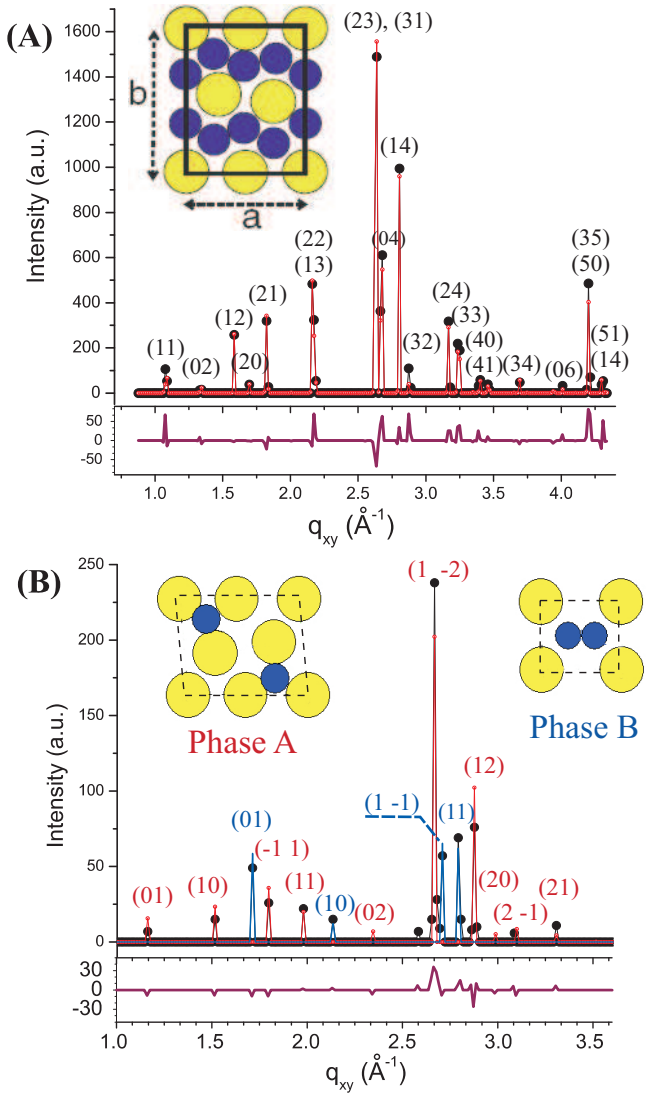}
\caption{Grazing incidence x-ray diffraction patterns with liquid bulk-like
contribution subtracted. The plots show the experimental data (circles), the theoretical fits (lines) and the difference (purple line, lower panels) for the
low-T (A)  and high-T (B) phases. Insets show the unit cells of
the two-dimensional crystalline phases corresponding to the fits.
The high-T diffraction patterns could only be fitted by a combination of
phases:  A (red peaks) and B (blue peaks).} \label{fig:ausi_fig5}
\end{figure}

Extensive attempts to index the high-$T$ GIXD pattern failed
to produce a single unit cell which reproduces all
observed peaks without producing additional strong peaks that were
not observed experimentally. However, in
Fig.~\ref{fig:ausi_fig5}(B) we show that the pattern is
reproduced well by two coexisting oblique two-dimensional
unit cells, shown in the inset. Phase A produces 10 GIXD peaks
shown in Fig.~\ref{fig:ausi_fig5}(B) in red. It has a unit
cell of $5.41 \times 4.25$~{\AA}$^2$ and
$\gamma=95.5\pm0.3^{\circ}$, containing 4 Au atoms and 2 Si atoms.
Phase B (4 GIXD peaks shown in blue) has a smaller unit cell of $3.66 \times
2.94$~{\AA}$^2$ and $\gamma=91.1\pm0.3^{\circ}$, containing 1 Au
atom and 2 Si atoms.

\begin{figure}[t]
\vspace{1mm} \centering
\includegraphics[angle=0,width=1.0\columnwidth]{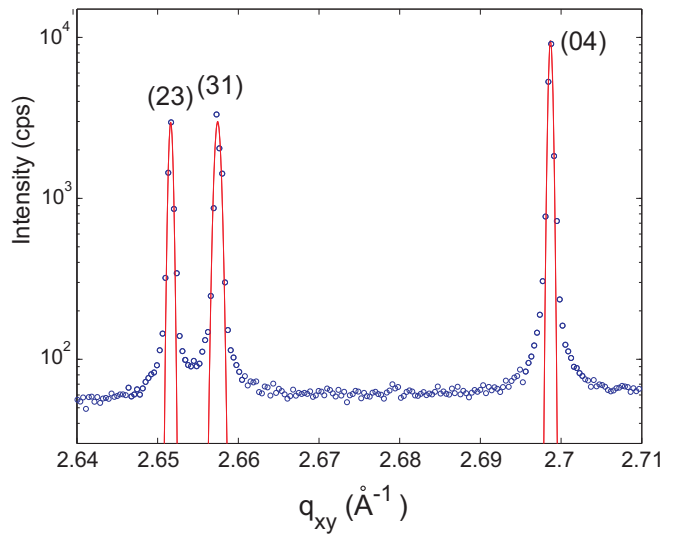}
\caption{Lineshapes of the (23), (31) and (04) GIXD peaks of the low-T phase measured with high-resolution Bonse-Hart Si (111) double-bounce crystal analyzer setup (blue circles). Fits with gaussian function (red lines) yields resolution-limited FWHM
of $\Delta q_{xy}\approx0.0005-0.0007$~{\AA}$^{-1}$. Peaks (23) and (31) appear as a single peak
with the low-resolution soller slit setup, but are well resolved in the pattern measured with the crystal analyzer.} \label{fig:ausi_fig6}
\end{figure}

The GIXD data discussed so far was collected with a
Soller slit analyzer in front of the detector, providing
an in-plane resolution of $\Delta
q_{xy}=0.011$~{\AA}$^{-1}$.With this resolution, the
widths of the diffraction peaks were resolution-limited.
The GIXD peaks were remeasured using a
double-bounce Bonse-Hart Si(111) analyzer,~\cite{Bonse}
having a resolution of $\Delta q_{xy} \approx
10^{-4}$~{\AA}$^{-1}$. High-resolution scans of selected
peaks are shown in Fig.~\ref{fig:ausi_fig6}.
Application of the Debye-Scherrer formula to the linewidths of low- and high-resolution GIXD lines, as well as stability of GIXD peak intensity during sample rotation imply micron-size crystallites.

The intensities of the GIXD peaks in the high-$T$ phase were found to fluctuate with
time. This could be assigned to fluctuations in the
orientation of the smaller number of larger (L$\gtrsim10~\mu$m)
crystallites in this phase. These fluctuations would then bring a
randomly varying number of crystallites into and out of the Bragg
condition, resulting in peak intensity fluctuations.

The thickness of the quasi-2D crystalline phase
can be estimated from the ratio of the integrated
intensities of the sharp diffraction peaks to that of the broad
liquid bulk peak. This ratio, 0.18, should be approximately equal
to the ratio of the thickness of the crystalline phase to the
decay length $\Lambda$ of the evanescent wave. For the
$\Lambda \approx 14$~{\AA} obtained above, this yields a
$\sim 2.5~${\AA} thick crystalline phase, or about a single atomic layer.

\begin{figure}[t]
\vspace{1mm} \centering
\includegraphics[angle=0,width=1.0\columnwidth]{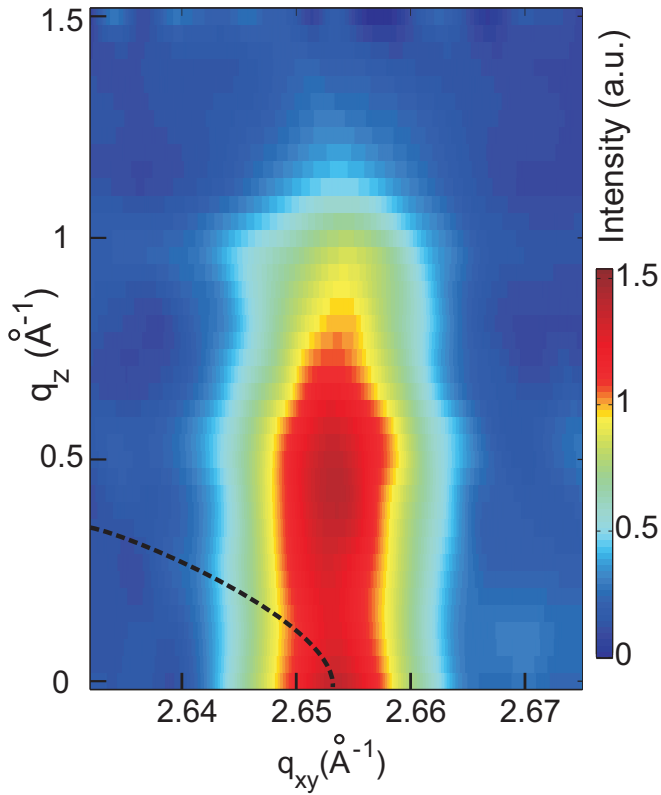}
\caption{Bragg rod trajectory
for (23) and (31) GIXD peaks centered at $q_{xy}=2.653$~{\AA}$^{-1}$ ,
collected with soller slits. Bragg rod is oriented along surface-normal direction $q_z$, indicative of a two-dimensional powder of crystallites aligned flat against the liquid surface. A
dashed line shows theoretically expected trajectory for a three-dimensional powder cone ($q_{xy}^2+q_{z}^2$=const).}\label{fig:ausi_fig7}
\end{figure}

\subsection{Bragg rod measurements}

A more precise estimate of the thickness of the crystalline
phase comes from a measurement of the Bragg
rods,~\cite{Robinson92} which are the intensity distributions
along $q_z$ at the GIXD peak positions.

The Bragg rod of the combined (23) and (31) GIXD peaks,
observed with the low-resolution setup at
$q_{xy}=2.653$~\AA$^{-1}$, was measured by employing a
fixed incidence angles $\alpha<\alpha_c$, and performing a series
of $2\theta$ scans at several (fixed) exit angles
$\beta$ relative to the surface.
As observed in
Fig.~\ref{fig:ausi_fig7}, the Bragg rod is oriented along
surface-normal ($q_z$) direction, indicating that it
originates in a 2D crystal and not in a 3D powder.
Had the GIXD peak originated in a 3D powder, a circular
"powder ring", shown in a dashed line in Fig.~\ref{fig:ausi_fig7},
would have been observed in the $(q_{xy},q_z)$ plane. This powder
ring is the locus of the Ewald sphere for which the
modulus of the scattering vector $\vec{q}$ is a constant value
$q=\sqrt{q_{xy}^2+q_z^2}$. Clearly, the measured rod does
not follow this trajectory, and thus does not originate in a 3D
powder.

\begin{figure}[b]
\vspace{1mm} \centering
\includegraphics[angle=0,width=1.0\columnwidth]{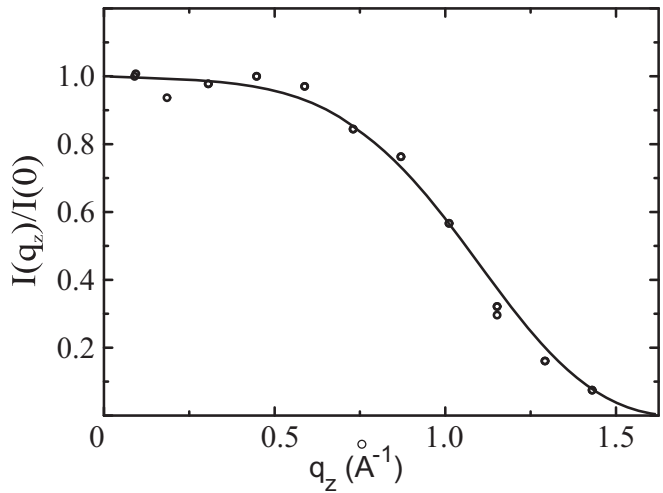}
\caption{Bragg rod intensity as a function of $q_z$, for the combined (23) and (31) GIXD peaks centered at $q_{xy}$=2.653~\AA$^{-1}$, measured with the low-resolution soller slits setup}
\label{fig:ausi_fig8}
\end{figure}

The intensity distribution along the Bragg rod ($q_z$ direction) is shown in
Fig.~\ref{fig:ausi_fig8} for the highest-intensity GIXD peak of the low-$T$ phase observed at low resolution at $q_{xy}=2.653$~\AA$^{-1}$.
Had this rod originated in a single, ideally flat planar monolayer the
intensity distribution along $q_z$ would have followed the form of the square of the atomic scattering factor of Au. In fact, the atomic scattering factor of Au does not vary by more than a few percent over the 1.5~\AA$^{-1}$ range
shown in Fig.~\ref{fig:ausi_fig8}. However, even for an
intrinsically sharp density profile, the surface roughening due to
capillary waves would have induced an exponential fall-off in the
Bragg rod intensity around $\sim1.5$~{\AA}$^{-1}$, consistent with
the fall-off observed in Fig.~\ref{fig:ausi_fig8}.
 It is also
possible that the fall-off near 1.5~{\AA}$^{-1}$ arises
from a combination of the capillary roughening and
a destructive interference between waves diffracted by the
2D order in two atomic layers. The condition for a
destructive interference of waves diffracted from two layers
separated by a distance $d$, $q_z\cdot d=\pi$, yields, in our case,
a distance  of $d=2.1$~{\AA}$^{-1}$. This is about 2/3 of the mean
layering distance (i.e. the distance between atomic layers measured
along surface normal) observed in the reflectivity measurements
discussed above. Moreover, this $d$ is also consistent with the
$\sim 2.5$~{\AA} thickness of the crystalline layer derived at the
end of the previous section from the intensity ratio of the sharp
GIXD peaks and the broad liquid peak of the bulk. A similar
estimation of the layer thickness from the intensity ratios for the
high-$T$ phase is complicated by the intensity fluctuations of the
GIXD peaks, mentioned above. Nevertheless, the crystalline layer
thickness of that phase seems to be comparable to that of the
low-$T$ phase. Clearly, Bragg rod measurements to larger $q_z$
values, and for additional GIXD peaks, are required to elucidate the
details of the 2D surface-parallel order in both low- and high-$T$
phases, and, in particular, their surface-normal variation.

\section{Discussion}

Detailed reasons for the formation of quasi-2D crystalline surface
layers in the eutectic AuSi alloy have not yet been established. Surface
melting (also known as pre-melting), i.e. a formation of a
quasi-liquid layer at the surface of a crystal below its bulk
melting point is a common and well-documented behavior for a wide
range of materials from metals and semiconductors to
dielectrics.\cite{Frenken85,Dash85} The existence of such a liquid-like
layer is the main reason for ice being slippery~\cite{Rosenberg05}
as has been first proposed in 1859 by Michael Faraday.\cite{Faraday}
Upon approaching the bulk melting point from below the crystalline structure
at the surface starts disordering because of the
increased entropy due to the lower number of near-neighbor atoms at
the surface. This coexistence of a disordered surface with an
ordered bulk is, thus, quite common. By contrast, the reverse effect
of the coexistence of a long-range ordered surface with a disordered
(liquid) bulk at a high temperature, as observed here, is quite
unusual.

Surface freezing, i.e. long-range crystalline in-plane ordering of
a surface layer in coexistence with an underlying liquid bulk, has been
previously reported in systems composed of long chain
molecules.\cite{Wu93, Dogic03} Several theoretical explanations have been suggested for
this effect, all of which (but one) are based on the highly
anisotropic shape of the molecules and their large,  $>20$~{\AA},
lengths.\cite{Tkachenko96} A different type of surface ordering effect
involves crystallization of the higher melting point dilute chemical
component in a binary metallic alloy, when it is segregated to the
surface, for example the crystallization of a surface-adsorbed monolayer
of Pb and of Tl in Ga$_{99.948}$Pb$_{0.052}$ \cite{Yang99} and
Ga$_{99.986}$Tl$_{0.014}$~\cite{Yang03} alloys, respectively. This
phenomenon differs substantially in several respects from the crystalline phase
reported here for Au$_{82}$Si$_{18}$ in several respects. As our (unpublished) studies of the freezing of a surface-segregated phase in a non-dilute alloy (GaBi)
show, the frozen layer (Bi in this case) is a macroscopically thick
crystalline surface film. This is not
surprising, since once the surface monolayer of pure Bi
crystallizes, it serves as a nucleating center for further crystal
growth. Second, the crystalline surface phase in eutectic AuSi is
not the pure chemical species, as is the case in GaPb and GaTl, and
in fact is not a phase that is stable in the bulk at any composition
or temperature. Finally, the solid-solid surface phase transition
observed in AuSi 12~$^\circ$C above the eutectic point, combined with
the stability of the high-temperature crystalline surface phase up to at
least 430~$^\circ$C, 70~$^\circ$C above the melting point, are difficult
to reconcile within the basic thermodynamics of the AuSi phase
diagram.

An intriguing question posed by this study is
the underlying reason for the formation of the crystalline monolayer
phase at the surface when such crystalline phase are not stable in the bulk. The unusually deep eutectic observed in AuSi, combined with the relative ease of formation
of a supercooled liquid and of glassy phases~\cite{Duwez60}, implies an increase resistance to, and a substantial degree of frustration against, crystal formation in this alloy. The existence of supercooling in liquid metals dates to observations by
Turnbull.\cite{Turnbull50} Based on these results, Frank
proposed~\cite{Frank52} the existence of a nucleation barrier due to
the icosahedral short-range order of the liquid. This prediction was recently confirmed  by
neutron~\cite{Schenk02} and x-ray~\cite{Kelton03} scattering studies
on electrostatically levitated supercooled metals and alloys.

Even though it is impossible to construct a long-range periodically ordered 3D bulk
phase based on close packing of polytetrahedral structures due to
the associated five-fold symmetry, this topological frustration may
be lifted at quasi-2D surfaces and interfaces. For example, x-ray
studies of liquid-solid interfaces have demonstrated that liquid
five-fold symmetry can be observed in monatomic liquid Pb when it is
aligned by the crystalline Si substrate.\cite{Reichert00} Packing
configuration consisting of two different kinds of atoms with
substantial size mismatch, as is the case of AuSi eutectic, results in a
far more complex topological problem compared to
monatomic liquids. However, one can draw an interesting parallel
between the low-temperature crystalline phase unit cell structure
shown in the inset of Fig.\ref{fig:ausi_fig5}(A) and the proposed
local structure in binary metallic glasses that is based on the idea of
overlapping, or interpenetrating clusters\cite{Miracle04,Sheng06} (compare the inset of Fig.\ref{fig:ausi_fig5} (A) with Fig. 3 in Ref. \cite{Miracle04}).
Because atoms within the surface layer have the additional freedom
of  moving slightly up or down (puckering, buckling or relaxation type effects), it is plausible
that an otherwise metastable 2D crystalline structure based on the
interpenetrating cluster model can be stabilized at the free
surface, even if long-range order cannot be extended into the 3D
bulk. Aside from these purely topological considerations, surface
and bulk electronic properties of the AuSi eutectic alloy are likely to
play a crucial role in the formation of the crystalline structure as
well. Changes in local chemical composition due
to Gibbs adsorption effects are expected to influence the short-range
order as well as the electronic behavior in the near surface region.
Amorphous AuSi alloys vary from semiconductors for Si-rich
compositions, characterized by continuous random network of
covalently bonded Si atoms, low packing density and a low atomic
coordination number (4 to 5), to metallic-like Au-rich AuSi glasses
of a high density due to random hard-sphere packing (high
coordination number of 8-9).\cite{Weis92} Since Si is expected to
have a lower surface tension than that of Au,\cite{Gibbs} the chemical
composition of liquid AuSi alloys evolves from Si-rich at the
surface to Au-rich in the bulk, resulting in a non-trivial evolution
of the electronic and the structural properties.\cite{Filonenko69}

Thin \emph{solid} films composed of metal-semiconductor alloys have been an
active area of research because of their interest for Schottky diode
devices.\cite{Xie88, Cros92} The crystalline surface phase reported
here does not resemble those of pure Au, pure Si or any of the
reported metastable bulk intermetalics.\cite{Okamoto83} Because of
the metastable nature of solid gold silicides, details of reported
intermetallic structures strongly depend on the methods of the film deposition
method, substrate surface type and preparation, annealing
temperature and duration, as well as other variables. An additional
difficulty is the accurate characterization of sub-nanometer thick
layers sandwiched between a solid substrate and a thick film of pure Au
or Si. However, crystalline phases with unit cell dimensions
$7.44 \times 9.33$~{\AA}$^2$, similar to our low-T phase, were
reported in thin Au films deposited on a Si(111)
surface.\cite{Green76, Green81} While no reliable thicknesses could be deduced for these phases from the low-energy electron diffraction
(LEED) and Auger electron spectroscopy techniques that were used,
sputtering studies indicate that the thickness of crystalline gold
silicide phases is in the range of 2-9~{\AA}.\cite{Green81}

Our investigations of other eutectic binary alloys involving Au and
elements of the IV semiconductor group, namely AuSn
\cite{Balagurusamy07} and AuGe (unpublished data) do not exhibit any
evidence for surface freezing effects akin to those reported here for AuSi. For these alloys the layering peak in the x-ray reflectivity signal is considerably weaker than that for AuSi,
and is consistent with a simple layering model comprising 3-4 atomic layers. Their GIXD shows no sharp diffraction peaks that would indicate
the existence of a long-range crystalline in-plane order in the near-surface region
of the liquid alloy. It is notable that the Au$_{72}$Ge$_{28}$ eutectic
shows nearly the same melting point ($\approx 360~^\circ$C) as the
Au$_{82}$Si$_{18}$ eutectic. Melting point of pure Ge,
938~$^\circ$C, is however much lower than that of Si
(1410~$^\circ$C). The size mismatch between Au and Si is much greater
than that between Au and Ge. While both AuGe and AuSi exhibit no stable
intermetalic compounds, Au and Si are not mutually soluble in solid
form, but solid Ge is weakly soluble in solid Au.\cite{Massalski}
These properties imply that the mismatch between atomic size is greater and the bonding of unlike species weaker in AuSi (evidenced by strong
covalent Si-Si bond~\cite{Chen67}) than in AuGe.

Further insight into the detailed reasons for the formation
of these unusual quasi-2D crystalline surface phases in coexistence
with the underlying liquid bulk might be ascertained from
theoretical and computer simulations of the evolution of the
electronic, chemical and structural properties of the near-surface
region of the AuSi eutectic, as well as future experimental
studies of the atomic structure of related liquid
metal alloys. Because of the close proximity of the eutectic
temperatures and Au compositions of AuSi and AuGe alloys, one of the
very promising approaches is the study of the evolution of surface
order in liquid ternary eutectic AuSiGe alloys
upon variation of Ge composition from zero up.

Finally, we note that the recent electron microscopy
studies of the atomic structure of liquid metals at their
interface with a solid\cite{Donnelly02, Oh05}, detailed
investigation of the chemistry of the
vapor-liquid-solid\cite{Hannon06} and the vapor-solid-solid
\cite{Kodambaka07} Si and Ge nanowire growth mechanisms, as well as
direct visualization of alternative crystallization pathways in
nanoscopic metallic droplets of AuGe alloys \cite{Sutter07} are related to the present study and our results might provide a better basis for understanding this rich phenomena.

\acknowledgements
This work was supported by the U.S. Department of
Energy (DOE) grant DE-FG02-88-ER45379 and the U.S.-Israel Binational
Science Foundation, Jerusalem. Work at Brookhaven National
Laboratory is supported by U.S. DOE contract DE-AC02-98CH10886.
ChemMatCARS Sector 15 is principally supported by NSF/DOE grant
CHE0087817. The Advanced Photon Source is supported by the U.S. DOE
contract W-31-109-Eng-38. Authors would like to acknowledge Jeff Gebhardt and
Frank Westferro for their contribution in making and testing of the
crystal analyzer. Discussions with Dr Stefan Mechler (Harvard)are gratefully acknowledged.

\bibliographystyle{unsrt}

\begin{thebibliography}{99}
\bibitem{Gibbs} Surface tension values of Au and Si at 370~$^{\circ}$C were extrapolated
to 1270-1300~mN/m for Au and 860-1000~mN/m for Si. Such large uncertainty is due to a large
temperature range over which extrapolation from melting points of pure metals was carried
out (700~$^{\circ}$C and 1030~$^{\circ}$C for Au and Si, respectively). Application of
Guggenheim equation, assuming interaction parameter set to zero results in 65-85 at. $\%$
of Si in the near-surface monolayer, in contrast with 19 at. $\%$ of Si in the bulk.
\end{thebibliography}

\end{document}